\documentclass[12pt,a4paper]{article}
\usepackage[utf8]{inputenc}
\usepackage[english]{babel}
\usepackage[T1]{fontenc}
\usepackage{natbib}
\usepackage{amsmath,amsthm}
\usepackage{amsfonts}
\usepackage{amssymb}
\usepackage{amsthm}
\usepackage{framed}
\usepackage{pdfpages}
\usepackage{tikz}

\usepackage{hyperref}
\newcommand{\imp}{\Rightarrow} 

\newtheorem{theorem}{Theorem}
\newtheorem{lemma}[theorem]{Lemma}
\newtheorem{corollary}[theorem]{Corollary}

\newtheorem{claim}[theorem]{Claim}
\newtheoremstyle{mydefinition}
  {1em}
  {1em}
  {\rmfamily}
  {0pt}
  {\bfseries}
  {: }
  {1pt}
  {\thmname{#1}\thmnumber{ #2}\thmnote{ (#3)}}

\theoremstyle{mydefinition}
\newtheorem{definition}{Definition}[section]

\begin{document}

\author{Jean-Gabriel Lauzier\footnote{We thank Nenad Kos and Massimo Marinacci for their support and Massimiliano Amarante, Ali Khan, Ekaterina Gavrilova, Christoph Wolf, Fabio Maccheroni, Philippe Grégoire and David Salib for their comments. We acknowledge financial support from Bocconi University.}\\
University of Waterloo}
\title{Ex-post moral hazard and manipulation-proof contracts
}
\maketitle
\begin{abstract}
We examine the trade-off between the provision of incentives to exert costly effort (ex-ante moral hazard) and the incentives needed to prevent the agent from manipulating the profit observed by the principal (ex-post moral hazard). Formally, we build a model of two-stage hidden actions where the agent can both influence the expected revenue of a business and manipulate its observed profit. We show that manipulation-proofness is sensitive to the interaction between the manipulation technology and the probability distribution of the stochastic output. The optimal contract is manipulation-proof whenever the manipulation technology is linear. However, a convex manipulation technology sometimes leads to contracts with manipulations in equilibrium. Whenever the distribution satisfies the monotone likelihood ratio property, we can always find a manipulation technology for which the optimal contract is not manipulation-proof.

\end{abstract}

\break
\section*{Introduction}
    
Ex-post moral hazard arguments have been widely used to rationalize features of real-world contracts. The earlier literature on financial contracts considers simple models where a borrower can lie about the real profit of a business while hiding money from the lender. Such manipulations provide a theoretical foundation for simple or collateralized debt contracts as optimal securities, as these contracts minimize the incentives to lie [\citep{attar2003costly}, \citep{lacker2001collateralized}]. Well-known macroeconomic models use similar arguments to microfound a borrowing constraint, as in \citet{kiyotaki1997credit}'s and \citet{bernanke1999financial}'s credit rationing.\\

Recent literature suggests that the rise of performance-based executive compensation is linked to an explosion of accounting scandals during the early twenty-first century, such as Enron's or Nortel Telecom's. Intuitively, the more CEOs are incentivized with bonuses, shares and options, the more incentives they have to use accounting techniques to make reported profits look higher than they are \citep{crocker2007economics}. In fact, the empirical literature on earnings management consistently observes a positive correlation between earnings management and CEOs' incentive pay. \citet{sun2014executive} and \citet{beyer2014optimal} suggest that this correlation may be driven by optimal contracting and is thus likely to be efficient. The idea that the optimal contract strictly trades-off between opposite incentives has also found ground in recent literature on securities design. The entrepreneurial financing model of \citet{koufopoulosoptimal} shows that bonus contracts, even while inducing manipulation in equilibrium, sometimes dominate debt contracts. Intuitively, debt contracts prevent manipulation perfectly while being incapable of separating entrepreneurial types when there is adverse selection. Bonus contracts do the exact opposite and are thus optimal when separating types is sufficiently valuable.\\

In other words, many strands of literature suggest that it is not always optimal to perfectly prevent the manipulation of observed profit. The implicit implication of such statement is that unethical behaviours like defrauding are to be expected in a well-functioning economy. That is, if such argument is to be true, then the unintended consequence of high-powered incentives is also to incentivize manipulation, and not much can or should be done to prevent this type of unethical behaviour.\\

To the best of our knowledge, no articles provide a set of general conditions for which the optimal contract entails manipulations in equilibrium. We are not aware of any general treatment which identifies the conditions under which the possibility of ex-post moral hazard is problematic or not. This situation is unfortunate given the strong normative implications of such models.\\

This paper aims to identify such set of general conditions. We build a general model of two-stage hidden actions and identify assumptions under which the optimal contract entails manipulation in equilibrium. To fix ideas, we interpret it as model of entrepreneurial financing where the entrepreneur can burn the business's money while having access to hidden borrowing.\footnote{The possibility of hidden borrowing is sometimes called window dressing.} We obtain two main results:\\

1) The optimal contract is manipulation-proof whenever the manipulation technology is linear. This holds for any distribution of profits. We can interpret the linearity of the manipulation technology as a situation where there are no frictions on the hidden borrowing market. This result implies that when the profit can take a continuum of values the optimal contract is a generalized debt contract with a bounded slope.\\

2) The optimal contract sometimes entail manipulations in equilibrium when the manipulation technology is convex. When the distribution of profit satisfies the monotone likelihood ratio property and another technical condition we can always find a manipulation technology for which the optimal contract is not manipulation-proof. The convex manipulation technology we consider in the main text can be interpreted as situations where there are frictions on the hidden borrowing market.\\

Intuitively, contracts with manipulation in equilibrium are optimal when they allow the entrepreneur to commit to a high(er) level of effort and the \textit{expected} waste is small. The manipulation technologies considered in the main text are particularly wasteful. During a manipulation the amount of resources wasted to manipulation can be very large with regard to the total profit made. However, such manipulations are infrequent when the effort is "productive enough", a difficult notion to pin down mathematically. We show that the monotone likelihood ratio implicitly makes the entrepreneurs' effort "very productive" and often leads to optimal contracts for which there is manipulation in equilibrium. This is an important observation because the monotone likelihood ratio is often assumed in application as a mathematical convenience to make the first-order approach valid. Our results thus imply that it is with loss of generality to ignore the possibility of manipulations in such applications.\\

There is no precise and agreed-upon definition of the term "manipulation" in the literature. We thus proceed with an extensive literature review that emphasizes the current paper's methodological innovation. We then present in Section 1 simple models with three states and two levels of efforts to provide intuition. But these simple models are not suited to stating or proving our results, and we develop in Section 2 the full model with a continuum of states and effort levels. Our second main result, the existence of contracts with manipulations in equilibrium, is obtained under the assumption of hidden borrowing. We discuss in Section 3 how this assumption effectively ties our hands, and explains why our second result is more general than it appears at first sight. In effect, relaxing the hidden borrowing assumption makes manipulations ever more acceptable. Appendix A is a primer on stochastic orders, Appendix B contains the proofs omitted in the text and Appendix C further explains the equilibrium concept we use.
\subsection*{Literature review}
    Many articles have examined the agency problem arising from an agent's ability to manipulate what is observed by the principal, and a precise definition of manipulation is essential to understand the literature. The current paper's narrative considers an entrepreneurial financing model where the entrepreneur can burn the enterprise's profit while having access to hidden borrowing and where the financier can observe the final profit at zero cost but can never observe the state. This type of manipulation is different from the situation where the entrepreneur sends a message about the state. When the entrepreneur can send any message at zero cost, he always has an incentive to declare a lower profit while keeping the money, essentially stealing from the business. This possibility leads to a complete failure of the lending market if no other mechanisms help mitigate the ex-post moral hazard problem. The costly state verification literature spawned by \citet{townsend1979optimal} assumes that the state is verifiable by the principal but at a high cost; see \citet{attar2003costly} for an excellent survey. \citet{lacker2001collateralized} also contains a similar manipulation technology but considers collateralization of the loan instead of verification of the state.\\

The possibility of stealing the money entails technical difficulties which are irrelevant to the point we want to make. We thus assume that the entrepreneur can burn the business' money, which implies that the optimal contract is non-decreasing. Our upward manipulation technology is formally close to the one found in the costly state falsification literature [\citep{lacker1989optimal}, \citep{maggi1995costly}, \citep{crocker1998honesty}], which assumes that the entrepreneur can send a message about the business' profit, that the state is not verifiable by the principal, but that lying is costly. The costly state falsification model of \citet{crocker2007economics} shows that perfectly preventing manipulation is prohibitively expensive in terms of opportunity cost when small lies are inexpensive. This is because their model's manipulation-proof contracts are entirely flat and do not incentivize work at all, and so the optimal contract strictly trades-off the provision of incentives and the prevention of manipulation. We show that the optimal contract sometimes entail manipulations in equilibrium even when the cost of manipulation is high. What matters is whether the expected gain of working harder outset the expected losses to manipulation. Our results thus nest \citet{crocker2007economics}'s findings, which assumes that the stochastic distribution of output satisfies the monotone likelihood ratio property.\\

Many articles investigate the link between the provision of incentives to CEOs and earnings management\footnote{The term earnings management refers to the possibility that a business' CEO can use accounting techniques to make a business' profit report appear better than it is.}. \citet{sun2014executive} argues that the positive empirical correlation between managers' incentive pay and earnings management is likely due to optimal contracting and does not necessarily reflect market inefficiencies. \citet{beyer2014optimal} analyses the optimal contract under earnings manipulation and relates the shape of the contract to the quality of the business' governance. This literature's manipulation technology is formally akin to the one found in the costly state falsification literature, and all comments we do regarding the latter literature applies to the former. Thus, we do not address directly the earnings management literature.\\

Our model can be interpreted as a problem of security design, with the caveat that we do not consider explicitly the competitive environment in which contracting happens. Our formalization of ex-ante moral hazard with limited liability borrows from \citet{innes1990limited}. \citet{koufopoulosoptimal}'s model of security design with both ex-ante hidden information (adverse selection) and ex-post moral hazard recently shed some new light on the role of bonus contracts. The authors ask whether the assumption that the returns to the lender are monotonic is without loss.\footnote{The article contains a thorough discussion of this monotonicity assumption in the security design literature.} They show that bonus contracts are sometimes optimal even though they fail this monotonicity assumption and induce manipulations in equilibrium. This is because bonus contracts allow for separating good and bad types of entrepreneurs, but contracts for which the lender's returns are monotonic do not. We show that this intuition carries to ex-ante hidden action, as bonus contracts provide high-powered incentives while keeping the expected waste to manipulation relatively low.\\

This article indirectly relates to the literature on the first-order approach's validity and the monotonicity of the optimal contract when the approach is valid. This literature shows how the monotone likelihood ratio property assumption (MLRP henceforth) is instrumental to modelling ex-ante moral hazard problems. This single assumption guarantees that both the underlying optimisation problem is easier to solve and that the optimal contract is monotonic. The subsequent literature made wide use of this assumption to simplify many applied problems. We refer to \citet{ke2018general} and \citet{ke2018monotonicity} for recent discussions. We show that if the distribution of profit satisfies the MLRP in effort then there are many situations for which the optimal contract entails manipulations in equilibrium. Thus, our results suggest that the MLRP is a stronger assumption than previously thought, as it precisely captures the notion that an "effort is productive enough" for the acceptance of such unethical behaviours in a well-functioning economy. \\

Modelling explicitly bonus contracts is challenging because it implies that the optimisation problem of the game's manipulation stage does not have well-defined first-order conditions. Fortunately, the game's manipulation stage is a positioning choice problem, a type of optimisation problem that we defined and characterized in \citet{Lauzier2019positioningmaths}. Positioning choice problems have a desirable property: their value function is always Lipschitz and almost everywhere differentiable. We extensively use this fact and other properties of positioning choice problems to simplify our proofs. However, we do not fully introduce the mathematical apparatus needed to solve positioning choice problems and refer the reader to the relevant theorems of \citet{Lauzier2019positioningmaths}.
\section{Motivating examples}
    We present simple models with three states and two levels of effort to provide the reader with intuition.\footnote{We need at least three states to make our point. With two states, the assumption that the distribution of profit satisfies first order stochastic dominance  in effort implies that it also satisfies the monotone likely ratio property. See appendix A for more details.} These examples are important because they illustrate well the mechanics underlying our main results. 

\subsection{Manipulation-proofness and linear technologies}
An entrepreneur needs to raise capital $Q> 0$ in order to finance a project.
The project profit is stochastic; let it be a discrete random variable taking
value $0\leq x_l <x_m<x_h=M$. The entrepreneur can take a costly action $e\in
E:=\{e_l,e_h\}$ that augments the expected profit of the project so that
$\mathbb{E}[X(e_h)]>\mathbb{E}[X(e_l)]$, where $X(e)$ is the stochastic profit
given effort level $e$. Exerting effort is costly, the cost $c$ of effort being
non-negative, increasing and convex, with $c(e_l)=0$. Further, we assume that
every random variable $X(e)$ has full support.\\

The hidden action $e\in E$ is chosen before the realization of the profit.
Then, in stage 2, the entrepreneur observes the realization $x\in \{x_l,
x_m,x_h\}$ of the profit and can  take a hidden action $z$. This
second hidden action modifies the profit $\overline{x}:=x+z$ by the
financier. For simplicity, we will assume throughout this section that the manipulation technology is restricted in such a way so that the observed profit always correspond to a realization so that $\overline{x}\in \{x_l,
x_m,x_h\}$. Finally, the financier observes $\overline{x}$ and the contract is
implemented without renegotiation.\\ 

The cost of the hidden action $z$ is parametrized by a function $g(z)$.
Consider for the moment that $g(z)=(1+r)max\{0,z\}$ for $r\geq 0$. We interpret
this function as the following manipulation technology: (A) when $z<0$ then
$g(z)=0$ and the entrepreneur burns the business' money and (B) the
entrepreneur can borrow the amount $z>0$ at the interest rate of $r$ and inject
the liquidities into the business, therefore inflating the business' observed profit.\\

Let us assume for simplicity that the financier's upfront payment is always $Q$. The contract
is therefore a vector $(y_i)_{i=l,m,h}$, where $y_i:=y(\overline{x}_i)$ is the
entrepreneur's state-contingent share of the profit upon the
financier observing $\overline{x}_i$. The financier keeps
$\overline{x}_i-y_i$. The entrepreneur is either risk-neutral or risk-averse, with standard Bernoulli
utility $u$ twice differentiable and weakly concave. The entrepreneur also has
both outside utility and limited liability normalized to zero. The financier is
risk-neutral with opportunity cost of investment $1+r$, where $r \geq 0$ is the
interest rate of the economy. Finally, let us also assume that the financier will never
pay more than the maximum profit realization of the project so that $y_i\leq
M$.\\

At time zero the financier makes a take-it-or-leave-it offer.\footnote{The
solution concept is a weak Perfect Bayesian Equilibrium where the entrepreneur
takes the action the most favoured by the financier whenever indifferent}
Denote by $\mathbb{P}[x_i \vert e]$ the conditional probability of $x_i$ given
effort level $e$. The financier's maximization problem is \begin{align*}
    \max_{(y(x_i + z_i))_{i=l,h,m},e\in E} &\sum_{i}\left(x_i +z_i - y(x_i + z_i)\right)\mathbb{P}[x_i \vert e] - Q \tag{Problem Discrete} \label{problem D}\\
    s.t.&\,   0\leq y(x_i + z_i) \tag{LL-D}\label{LLD}\\
     & y(x_i+ z_i)\leq M \tag{B-D}\label{BD}\\
     & \, \sum_{i} u\left(y(x_i+z_i) - g(z_i) \right)\mathbb{P}[x_i\vert e]  -c(e) \geq 0 \tag{IR-E-D}\label{IRED}\\
     &\sum_{i}\left(x_i +z_i - y(x_i + z_i)\right)\mathbb{P}[x_i \vert e] \geq (1+r) Q \tag{IR-F-D} \label{IRFD}\\
    & \, e\in arg\max_{\hat{e}\in E} \left\{ \sum_{i} u\left(y(x_i+z_i) - g(z_i)  \right) \mathbb{P}[x_i \vert \hat{e}] -c(\hat{e}) \right \} \tag{IC-D} \label{ICD}\\
    & \, \forall x_i,\, z_i \in arg\max_{z} \{ y(x_i+z) - g(z)\}\tag{IIC-D} \label{IICD}
\end{align*}
where \eqref{LLD} is the limited liability constraint, \eqref{BD} is the
boundedness constraint, \eqref{IRED} is the individual rationality constraint
of the entrepreneur, \eqref{ICD} is the incentive compatibility constraint
defined by stage 2 and \eqref{IICD} is the interim incentive compatibility
constraint defined by stage 3. Since we are maximizing the financier expected profit, we ignore the constraint \eqref{IRFD} while solving and simply verify that it is not binding. \\

Consider now the distribution in Table 1.
\begin{table}[!ht]
\centering
\caption{A distribution satisfying FOSD but not MLRP}
\begin{tabular}{c|ccc}
                        
   &    $\mathbb{P}[x_l \vert e]$ & $\mathbb{P}[x_m \vert e]$& $\mathbb{P}[x_h \vert e]$\\ \hline
$e_l$ & 0.5              & 0.49995             &  0.00005             \\
$e_h$ & 0.5             & 0.00005             & 0.49995             
\end{tabular}
\end{table}
This distribution satisfies the assumption of first-order stochastic dominance
(FOSD) in effort but does not satisfy the monotone likelihood ratio property
(MLRP). Absent ex-post moral hazard if the entrepreneur is risk-averse and if
the financier wants to implement the high level of effort $e_h$ then the
optimal contract is non-monotonic, i.e. $y_m<y_l<y_h$. However, this contract
is not optimal if we consider the possibility of manipulations.

\subsubsection*{Property 1 - monotonicity}
We show that the possibility of burning the money implies that the payoff of the entrepreneur must be monotonic. Suppose that the optimal contract with ex-ante moral hazard only is not monotonic, so that $y_m<y_l<y_h$, and that it satisfies the following inequality:
$$y_l > y_h - (1+r)(x_h-x_m).$$  This contract
is not optimal if the entrepreneur can manipulate the observed profit.\\

In fact, the inequality implies that the entrepreneur would like to manipulate downward; upon realization of profit $x_m$, the entrepreneur burns the amount $x_m-x_l$
and receives $y_l>y_m$. Consider now an alternative contract where $$y'_m=y_l=y'_l \quad \text{
and} \quad y'_h=y_h.$$ This new contract strictly dominates the original contract
as it does not induce wasteful manipulations and does not change the
incentives to exert effort, as the entrepreneur receives state-by-state the same
amount with both contracts. In other words, the value function of the optimisation problem defined at the
manipulation stage of the game is non-decreasing whenever the entrepreneur can
freely burn money. This implies that any contract which is decreasing
somewhere is dominated by a monotonic contract, since replacing the former
by its monotone envelope does not change incentives.\footnote{See section 2.1. for a definition of the monotone envelope (definition 2.1).}
    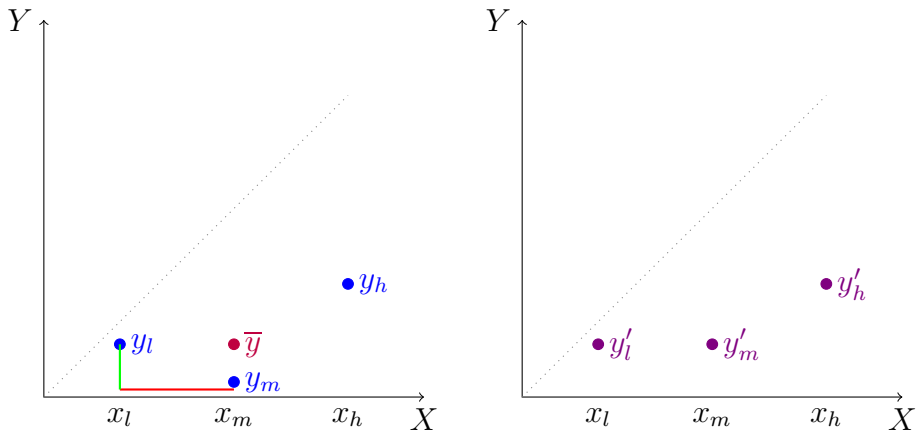
\begin{figure}[!ht]
    \centering
    \caption{Monotonicity}
    $\begin{array}{lr}

\begin{tikzpicture}
\draw[<->] (0,5) -- (0,0) -- (5,0);
\draw[gray,dotted] (0,0) -- (4,4);
\node[below] at (5,0) {$X$};
\node[left] at (0,5) {$Y$};
\node[below] at (1,0) {$x_l$};
\node[below] at (2.5,0) {$x_m$};
\node[below] at (4,0) {$x_h$};
\filldraw[blue] (1,0.7) circle (2pt) node[anchor=west] {$y_l$};
\filldraw[blue] (2.5, 0.2) circle (2pt) node[anchor=west] {$y_m$};
\filldraw[blue] (4, 1.5) circle (2pt) node[anchor=west] {$y_h$};
\filldraw[purple] (2.5, 0.7) circle (2pt) node[anchor=west] {$\overline{y}$};
\draw[red,thick] (1,0.1) -- (2.5,0.1);
\draw[green,thick] (1,0.1) -- (1,0.7);
\end{tikzpicture}  

&
\begin{tikzpicture}
\draw[<->] (0,5) -- (0,0) -- (5,0);
\draw[gray,dotted] (0,0) -- (4,4);
\node[below] at (5,0) {$X$};
\node[left] at (0,5) {$Y$};
\node[below] at (1,0) {$x_l$};
\node[below] at (2.5,0) {$x_m$};
\node[below] at (4,0) {$x_h$};
\filldraw[violet] (1,0.7) circle (2pt) node[anchor=west] {$y_l'$};
\filldraw[violet] (4, 1.5) circle (2pt) node[anchor=west] {$y_h'$};
\filldraw[violet] (2.5, 0.7) circle (2pt) node[anchor=west] {$y_m'$};
\end{tikzpicture}  
\end{array}$

\end{figure}

\subsubsection*{Property 2 - bounded slope} We now want to show that a linear upward manipulation technology implies that the slope of the optimal contract must be bounded. Suppose again that the contract with ex-ante moral hazard only is non-monotonic, $y_m<y_l<y_h$, but that it has a large spread in payoff between the mid and high states. Formally, let the contract with ex-ante moral hazard satisfies the following inequality: 
\begin{align*}
y_l < y_h - (1+r)(x_h-x_m).
\end{align*} 
Clearly, upon the mid realization $x_m$ the entrepreneur will inject the amount $x_h-x_m$ in the business, hitherto declaring $\overline{x}_h$. The effective payoff upon realization $x_m$ is therefore $$\overline{y}_m=y_h-(1+r)(x_h-x_m).$$
Again, this contract cannot be optimal if the entrepreneur can manipulate the observed profit. Updating the contract with the manipulation
stage's value function, i.e. replacing $y_m$ with $\overline{y}_m$, eliminates wasteful manipulations while leaving incentives intact.\\

This argument is always true when $g(z)$ is linear, which implies that the optimal contract has a bounded slope. Formally, we prove this statement by showing that we can always replace a contract by the value function of the optimisation problem it defines in the manipulation stage without changing the incentives to work. The value function always has a bounded slope and so must the optimal  contract. However, this is only true for linear manipulation technologies, as it is the only type of technology for which the entrepreneur receives state-by-state the same amount with both contracts.

    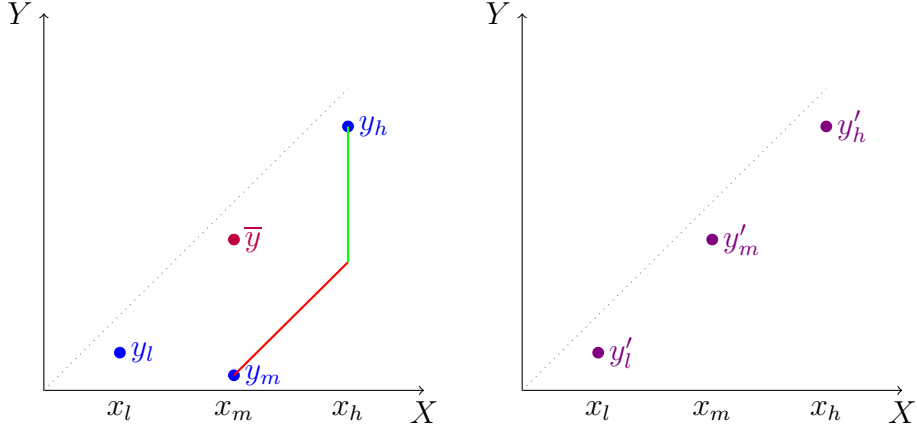
\begin{figure}[!ht]
    \centering
    \caption{Linear manipulation technology and bounded slope}
    $\begin{array}{lr}

\begin{tikzpicture}
\draw[<->] (0,5) -- (0,0) -- (5,0);
\draw[gray,dotted] (0,0) -- (4,4);
\node[below] at (5,0) {$X$};
\node[left] at (0,5) {$Y$};
\node[below] at (1,0) {$x_l$};
\node[below] at (2.5,0) {$x_m$};
\node[below] at (4,0) {$x_h$};
\filldraw[blue] (1,0.5) circle (2pt) node[anchor=west] {$y_l$};
\filldraw[blue] (2.5, 0.2) circle (2pt) node[anchor=west] {$y_m$};
\filldraw[blue] (4, 3.5) circle (2pt) node[anchor=west] {$y_h$};
\filldraw[purple] (2.5, 2) circle (2pt) node[anchor=west] {$\overline{y}$};
\draw[red,thick] (2.5,0.2) -- (4, 1.7);
\draw[green,thick] (4,1.7) -- (4,3.5);
\end{tikzpicture}  

&
\begin{tikzpicture}
\draw[<->] (0,5) -- (0,0) -- (5,0);
\draw[gray,dotted] (0,0) -- (4,4);
\node[below] at (5,0) {$X$};
\node[left] at (0,5) {$Y$};
\node[below] at (1,0) {$x_l$};
\node[below] at (2.5,0) {$x_m$};
\node[below] at (4,0) {$x_h$};
\filldraw[violet] (1,0.5) circle (2pt) node[anchor=west] {$y_l'$};
\filldraw[violet] (4, 3.5) circle (2pt) node[anchor=west] {$y_h'$};
\filldraw[violet] (2.5, 2) circle (2pt) node[anchor=west] {$y_m'$};
\end{tikzpicture}  
\end{array}$
\end{figure}

\subsection{Acceptable manipulations with convex technologies}

Next we present the intuition behind the optimality of contracts with manipulation. Let 
\begin{align*}
    g(z) = \begin{cases}0 &\text{ if } z\leq 0 \\ \Tilde{g}(z) &\text{ if } z>0 \end{cases}
\end{align*}
for $\Tilde{g}$ a strictly convex function satisfying 
\begin{align*}
    \lim_{z\downarrow 0}\Tilde{g}(z)=0 \quad \text{and} \quad \inf_{z>0} \left \{\frac{\partial \Tilde{g}(z)}{\partial z}\right \} \geq 1.
\end{align*}
This assumption guarantees that $g$ is continuous and that lying is always "expensive". We interpret this manipulation technology as a situation where there are frictions on the hidden borrowing market so that the interest rate is increasing in the size of the loan.\\

Let us assume that the entrepreneur makes the take-it or leave-it offer at the initial stage of the game. Let $q\in (0,0.45)$ and consider the distribution in Table 2. We will show that when the conditional probability of the middle state $x_m$ vanishes, $q\downarrow 0$, the entrepreneur is better-off proposing a contract that entails manipulation in equilibrium.
\begin{table}[h!]
\centering
\caption{Another distribution that satisfies FOSD but not the MLRP}
\begin{tabular}{c|ccc}
                        
   &    $\mathbb{P}[x_l \vert e]$ & $\mathbb{P}[x_m \vert e]$& $\mathbb{P}[x_h \vert e]$\\ \hline
$e_l$ & 0.1  & 0.9 - q     &  q             \\
$e_h$  & 0.1 & q & 0.9 - q          
\end{tabular}
\end{table}

\subsubsection*{Best manipulation-proof contract}

Let us first consider a manipulation-proof contract $Y=\{y_l,y_m, y_h\}$. The possibility of manipulation defines the following upward interim incentive compatibility constraints:
\begin{align}
  y_m -g(x_m-x_l) \leq y_l \tag{$ IIC_{l,m}$} \label{IIClm}\\
     y_h -g(x_h-x_m) \leq y_m \tag{$ IIC_{m,h}$} \label{IICmh}\\
   \quad y_h -g(x_h-x_l) \leq y_l \tag{$ IIC_{l,h}$} \label{IIClh}.
\end{align}
Consider now the manipulation-proof contract that maximizes the spread between the low and high states. This means that the constraints $IIC_{l,m}$ and $IIC_{m,h}$ are binding, i.e.
\begin{align*}
    IIC_{l,m}:& \quad y_m -g(x_m-x_l)  =  y_l \\
    IIC_{m,h}:& \quad y_h -g(x_h-x_m) = y_m. 
\end{align*}
This is the manipulation-proof contract that maximizes the incentives to exert high effort. It is important to notice that the constraint \eqref{IIClh} is slack because $g$ is convex. The previous equalities can be rewritten as
\begin{align*}
 IIC_{l,m}:& \quad  y_m =  y_l + g(x_m-x_l)\\
IIC_{m,h}:& \quad y_h = y_l + g(x_m-x_l)+g(x_h-x_m)
\end{align*}
and thus for $y_l$ and $e$ given the expected payoff  $\mathbb{E}[u(Y)\vert e]$ is
\begin{align*}
    \mathbb{P}[x_l\vert e]u(y_l) + \mathbb{P}[x_m\vert e]u(y_l+ g(x_m-x_l))+ \mathbb{P}[x_h\vert e]u( y_l + g(x_m-x_l)+g(x_h-x_m)).
\end{align*}

\subsubsection*{Contract with manipulation}
Let us now consider a contract that entails manipulation in equilibrium, but only for the realization $x_m$. Let $Y^{M}=\{y^{M}_l,y^{M}_m, y^{M}_h\}$ be a contract for which $IIC_{l,h}$ is an equality, i.e $$y_h^M = y_l^M+ g(x_h-x_l).$$ By the strict convexity of $g$ we have $$y_h^{M}=g(x_h-x_l) +y^{M}_l> g(x_h+x_m) +g(x_m-x_l) +y^{M}_l$$ which implies that $IIC_{m,h}$ is violated whenever $IIC_{l,m}$ is satisfied. This means that upon the realization $x_m$, the entrepreneur always injects $x_h-x_m$ in the business and declares $\overline{x}_h$. Proceeding as before, the expected payoff $\mathbb{E}[u(Y^M)\vert e]$ for $y_l^M$ and $e$ given is 
\begin{align*}
    \mathbb{P}[x_l\vert e]u(y_l^M) + \mathbb{P}[x_m\vert e]u(y_l^M+g(x_h-x_l) - g(x_m-x_l))+ \mathbb{P}[x_h\vert e]u(y_l^M + g(x_h-x_l)).
\end{align*}
\subsubsection*{The optimal contract sometimes entails manipulations}
We just showed that the interim incentive compatibility constraints entirely pin down the spread in payoff between the low and high realizations. Now observe that for $e$ given, the entrepreneur wants to set $y_l$ and $y_l^M$ as high as possible. However, it is easily shown that $y_l\geq y_l^M$ for any given $e$, which means that the manipulation-proof contracts give higher payoff in lower states. This implies that contracts with manipulations sometimes implement strictly higher levels of effort than the best manipulation-proof contract. In our example, the value of exerting effort increases as the probability of the middle state $q$ vanishes. In other words, when $q\downarrow 0$, there exists situations where the optimal contract entails manipulation in equilibrium.

\section{Full model}
    
We now present the full fledged model with a continuum of states and effort
levels to prove our main results. We show in Section 2.1 that the optimal
contract is non-decreasing whenever the entrepreneur can burn the business'
money. We also state in this section two lemmas inspired by \cite{Lauzier2019positioningmaths} which help solve the full fledged
model. We show in Section 2.2 that manipulation-proofness obtains whenever the
manipulation technology is linear. As a corollary, we obtain that the optimal
contract is a generalized debt contract with a bounded slope. Finally, we show
in section 2.3 that when the distribution of profit satisfies the MLRP and
another technical condition then we can always find a convex manipulation
technology for which the optimal contract entails manipulation in equilibrium.
\\

Let the set of efforts be $E=[0,e_{max}]$ for $e_{max}>0$ large. The business'
profit is a family of continuous random variables $(X(e))_{e\in E}$ with common
and full support $[0,M]$. Exerting effort augments the expected profit of the
project so that $e>e'$ implies $\mathbb{E}[X(e)]>\mathbb{E}[X(e')]$.\footnote{The vector
notation $(\cdot)$ is used instead of the general $\{\cdot\}$ to emphasize that
we consider families of random variables ordered in one of
the stochastic orders defined in Appendix A.} Exerting
effort is costly, with the cost $c:E\rightarrow\mathbb{R}_+$ being increasing,
differentiable, (weakly) convex and satisfying $c(0)=0$. We introduce states of the worlds to keep the notation compact. Let $S=[0,M]$ be a set of states of the world and let $\mathcal{B}(S)$ be the
Borel sigma-algebra of $S$. The family of random variables defined above is
thus a family of $X: S \times E  \rightarrow [0,M]$ for which it is assumed that for every
$e\in E$, $X(s,e)$ is a continuous and weakly increasing function of $s$ such that $$\min_{s\in S}X(s,e)=0<M:=\max_{s\in S} X(s,e)<+\infty.$$ 

The entrepreneur needs to borrow the amount $Q>0$ before starting the project.
Let us assume again that the financier is risk-neutral and never pays more than
$Q$. The contract is a transfer function $$Y(X) \in
B_+(\mathcal{B}([0,M]))$$ which depends only on the observed realization of
profit $\overline{x}(s)$, where $B_+(\mathcal{B}([0,M]))$ denotes the Banach
space (sup-norm) of non-negative bounded functions which are measurable with
regard to $\mathcal{B}([0,M])$. This function $Y$ represents the amount received by the
entrepreneur, with the financier keeping the amount $\overline{x}-Y(\overline{x})$. The entrepreneur first chooses the level of effort and then observes the state
$s\in S$. The entrepreneur can then take another hidden action $z$ in
order to manipulate the profit $\overline{x}=X(s,e)+z$ observed by the
financier.\footnote{We assume that the feasible manipulation $z$ is contingent on the state and is such that $\overline{x}=x+z\in [0,M]$ so that the observed profit always correspond to a possible realization of the random profit $X(s,e)$. However, keeping track explicitly of this constraint is very cumbersome, and we will not specify it.} The cost of this hidden action is parametrized by a function
$g:[-M,M]\rightarrow \mathbb{R}_+$ which represents the manipulation
technology. We assume that \begin{align*}
    g(z) = \begin{cases}0 &\text{ if } z\leq 0 \\ \Tilde{g}(z) &\text{ if } z>0 \end{cases}
\end{align*}
for $\Tilde{g}$ a (weakly) convex function that is differentiable on $(0,M)$ and that satisfies
\begin{align}
    \lim_{z\downarrow 0}\Tilde{g}(z)=0 \quad \text{and} \quad \inf_{z>0} \left \{\frac{\partial \Tilde{g}(z)}{\partial z}\right \} \geq 1.\tag{Assumption 1} \label{Assumption 1}
\end{align}
The first part of \ref{Assumption 1} is to guarantee that the cost of a
manipulation is a continuous function, while the second is to guarantee that
inflating the observed profit is always expensive. We interpret this manipulation technology as a situation where reducing the profit is costless, but where inflating requires for the entrepreneur to borrow money. We further discuss the
interpretation of this manipulation technology in Section 3.\\

The entrepreneur is either risk-neutral or risk-averse, with Bernoulli utility
$u$ weakly concave and differentiable. The entrepreneur has outside
utility $\overline{u}\geq 0$ and limited liability standardized to zero so that
$Y\geq 0$. Similarly, we assume that $Y\leq M$. This boundedness
constraint states that the financier never pays the entrepreneur more than the
maximum amount which the business can make. The financier makes a take-it or leave-it offer at the initial stage of the
game. The solution concept is a weak Perfect Bayesian equilibrium where the
entrepreneur takes the action most favoured by the financier when
indifferent.\footnote{We consider weak Perfect Bayesian Equilibria in pure
strategies \citep{mas1995microeconomic} with the assumption that the
entrepreneur chooses the highest level of effort whenever indifferent and takes
the manipulation most favoured by the financier whenever indifferent. See Appendix C
for more details.} By backward induction the optimal contract satisfies the
optimisation program 
\begin{align}
\sup_{Y \in B_+(\mathcal{B}([0,M])), \overline{e} \in E} &\int X(s,\overline{e}) + z(s) - Y(X(s,\overline{e})+z(s)) d \mathbb{P} - Q \tag{Problem F} \label{Problem F}\\
  s.t.\, & \,0 \leq Y \tag{LL} \label{LL}\\
       & \, Y \leq M \tag{B} \label{B}\\
       & \, \int u\left (Y(X(s,\overline{e}) +z(s)) -g(z(s)) -c(\overline{e})\right )d\mathbb{P} \geq \overline{u} \tag{E-IR} \label{E-IR}\\
       &\, \int X(s,\overline{e})+ z(s) - Y(X(s,\overline{e})+z(s))) d \mathbb{P} \geq (1+r)Q  \tag{F-IR} \label{F-IR}\\
       &\, \overline{e}\in \arg \max_e \left \{\int u(Y(X(s,e) +z(s)) -g(z(s)) -c(e))d\mathbb{P} \right \}\tag{IC} \label{IC}\\
       \forall s \in S,\, &\quad z(s) \in  \arg \max_z \{Y(X(s,\overline{e})+ z) -g(z) \} \tag{IIC} \label{IIC}
\end{align}
where \eqref{E-IR} is the entrepreneur's participation constraint, \eqref{F-IR}
is the financier's participation constraint, \eqref{IC} is the incentive
compatibility constraint imposed by stage one and \eqref{IIC} is the interim
incentive compatibility constraint imposed by stage two. Since we are maximising the financier's payoff, we will ignore the constraint \ref{F-IR} while solving the problem and verify that the solution we have found satisfies it.

\subsection{Monotonicity of the optimal contract}

Each contract $Y$ defines a sequential choice of effort $e$ and then of
manipulation $z$. At the manipulation stage both $e$ and $s$ are given so we
can write the \textit{optimal choice correspondence} of this stage as
\begin{align*}
   \sigma(Y,x_{s,e}) :=\arg\max_{z} \{Y(X(s,e)+ z) -g(z) \}.
\end{align*}
The \textit{value function} of the manipulation stage of the game is
\begin{align*}
    V(x_{s,e};Y) := Y(X(s,e)+z(s))-g(z(s))
\end{align*}
for $z(s)$ a selection of
$\sigma(Y,x_{s,e})$.\footnote{A selection $f$ of a correspondence $F$ is a
function such that for every $x\in domain(F)$ it is $f(x)\in F(x)$.} Intuitively, we want to allow $Y$ being discontinuous because
we interpret the upward jumps as bonuses. However, this makes characterizing
the optimal choice correspondence and the value function much harder. This is because we cannot take first-order conditions to pin-down the optimal manipulations as the manipulation stage's objective function is discontinuous and thus not differentiable.\\

Fortunately, the optimisation problem of the manipulation stage is a
positioning choice problem, a class of optimisation problems which we defined and
examined in \citet{Lauzier2019positioningmaths}. This observation will help us simplify the treatment. Without loss of generality we consider contracts that are
almost everywhere continuous and that satisfy the following technical
assumption: \begin{align}
    \text{ for every }\,x\in[0,M] \text{ it is } \lim\sup_{x'\rightarrow x}Y(x')=Y(x). \tag{Assumption 2} \label{Assumption 2}
\end{align}
We can now state two lemmas which help solve problem \ref{Problem F}. \\

\begin{lemma}[Monotonicity of the value function]\label{monotonicityVF}
Let the function $g$ satisfy \ref{Assumption 1}. Then for every given $Y$ and $e$ the value function $V(x_{s,e};Y)$
is non-decreasing.
\end{lemma}

\noindent \textbf{Proof of Lemma 1:} 
The statement is trivial if $Y$ is non-decreasing. Suppose that $Y$ is decreasing somewhere. Then to every $x$ in a decreasing segment of $Y$ there exists a $x'<x$ such that for $z=-(x-x')<0$ it is
$$Y(x)<Y(x+z) - g(z)= Y(x')$$
and $V(x_{s,e},Y)$ is non-decreasing. $\blacksquare$\\

Lemma 1 tells us that the entrepreneur can always make it look like if he made less profit, and thus receive the higher payoff. The next lemma will also be useful.

\begin{lemma}[Continuity of the value function]\label{Continuity}
Let the function $g$ satisfy \ref{Assumption 1}. Then for every given $Y$ and $e$ the value function $V(x_{s,e};Y)$
is Lipschitz continuous and thus almost everywhere differentiable.
\end{lemma}

Lemma 2 tells us that no matter the shape of the contract $Y$, the value function of the manipulation stage of the game is \textit{always} a continuous function. This observation helps us in our search for an optimal contract. We refer the reader to Appendix B for the proof. We are now ready to prove our first statement. 

\begin{theorem}[Monotonicity of the optimal contract]\label{MonotonicityY}
Any contract solving \ref{Problem F} is non-decreasing.
\end{theorem}
The proof is instructive and will be done carefully. It uses the following:
\begin{definition}[Monotone envelope]
Let $Y$ satisfy \ref{Assumption 2}. The \textit{monotone envelope} of the
function $Y$ is the smallest non-decreasing function $\overline{Y}$ such that
$Y\leq \overline{Y}$.\\ \end{definition}
\noindent \textbf{Proof of Theorem \ref{MonotonicityY}:} Suppose by
contraposition that the contract $Y$ is decreasing somewhere and let $e_Y$ be a
level of effort that contract $Y$ implements. By Lemma
\ref{monotonicityVF} and \ref{Continuity} the value function \begin{align*}
    V(x_{s,e_Y};Y)
\end{align*}
is continuous and non-decreasing. By \ref{Assumption 1} it is also the case
that $V(x_{s,e_Y};Y)\geq Y$. Consider the alternative contract $\overline{Y}$
defined by the monotone envelope of $Y$.\\
\begin{lemma}\label{Monotoneenvelope}
The contract $\overline{Y}$ implements $e_Y$ and is such that 
\begin{align*}
    V(x_{s,e_Y};Y)=V(x_{s,e_Y};\overline{Y}).
\end{align*}
\end{lemma} 
\noindent \textbf{Proof of Lemma \ref{Monotoneenvelope}:} Consider $x\in [0,M]$ and redefine the choice correspondence as $$\sigma(Y,x)=\arg\max_{z}\{Y(x+z)-g(z)\}$$ and the value function as
$$V(x;Y)= Y(x+z(x))-g(z(x))$$
for $z(x)\in \sigma(Y,x)$.
By definition if $0\in \sigma (Y,x)$ then $0\in \sigma(\overline{Y},x)$ and $V(x;Y)=V(x;\overline{Y})$. It remains to show the cases when $0\notin \sigma (Y,x)$.\\

\noindent\textit{Downward manipulation:} If there exists a $z \in\sigma(Y,x)$ such that $z<0$ then $$V(x;Y)=Y(x+z)-g(z)= Y(x+z)=\overline{Y}(x)$$
by definition of the monotone envelope. Thus, $0\in\sigma(\overline{Y},x)$ and $V(x;Y)=V(x;\overline{Y})$.\\

\noindent \textit{Upward manipulation:} If every $z\in \sigma(Y,x)$ are such that $z>0$ then $\sigma(Y,x)=\sigma(\overline{Y},x)$ and $V(x;Y)=V(x;\overline{Y})$.\\

We have just shown that for every given level of effort the value functions of the
manipulation stage of the game are equal under both contracts. Thus,
$\overline{Y}$ implements effort $e_Y$ .$\square$\\

By construction the contract $\overline{Y}$ satisfies constraint \eqref{E-IR} if the contract $Y$ does. Let $z(s)\in \sigma (Y,x_{s,e_Y})$ and 
$\overline{z}(s)\in\sigma(\overline{Y},x_{s,e_Y})$ be selections. The contract $\overline{Y}$ dominates the contract $Y$ since the
latter induces downward manipulation which implies that \begin{align*}
    \int X(s,e_Y) + \overline{z}(s) - \overline{Y}(X(s,e_Y)+\overline{z}(s)) d \mathbb{P} > \int X(s,e_Y) + z(s) - Y(X(s,e_Y)+z(s)) d \mathbb{P},
\end{align*}
 and $Y$ is not optimal. $\blacksquare$\\

The critical steps are in Lemma \ref{Monotoneenvelope}. Virtually all this article's
proofs rely on comparing two contracts and verifying whether or not they
implement the same level of effort. Theorem \ref{MonotonicityY} compares a
contract to its monotone envelope because Lemma \ref{monotonicityVF} implicitly
guarantees that they implement the same level of effort. This monotonicity property is entirely driven by the manipulation technology
and does not rely on properties of the distribution of profit. We now show that
a similar result is true for linear manipulation technologies.

\subsection{Linear manipulation technologies and manipulation-proofness}

Let us now assume that 
\begin{align}
    g(z)=(1+r)\max\{0,z\} \quad \text{for} \quad r\geq 0 \tag{Assumption 3}. \label{assumption3}
\end{align} 
The following ancillary lemma immediately obtains:\\
\begin{lemma}\label{boundedslopeVF}
Let $g$ satisfy \ref{assumption3}. For every given $Y$ and $e$ the value function $V(x_{s,e};Y)$
has a slope lesser or equal to $1+r$.
\end{lemma}

Lemma \ref{boundedslopeVF} allows us to essentially repeat the proof of
Theorem \ref{MonotonicityY} while using the value function defined by a
contract $Y$ instead of its monotone envelope.
\begin{theorem}[Manipulation-proof contracts]\label{MPtheorem}
Let $g$ satisfy \ref{assumption3}. Then the optimal contract $Y$ is
manipulation-proof: for every $x\in [0,M]$ it holds that
$$0\in \arg\max_{z}\{Y(x+z)-g(z)\}.$$ 
Moreover, it is continuous, non-decreasing and has a slope lesser or equal to $1+r$, i.e.
\begin{align*}
    0\leq \frac{\partial Y(x)}{\partial x}\leq 1+r.
\end{align*}
whenever this derivative is well-defined.
\end{theorem}

As mentioned, the proof of Theorem \ref{MPtheorem} is almost identical to the
proof of Theorem \ref{MonotonicityY}. For the sake of brevity we omit it in the
main text and refer the reader to Appendix B. It is worth emphasizing again that the proof does not rely on properties of the
distribution of profit. The linearity of the manipulation technology entirely
drives the result because this is what allows us to replace any contract by the
value function of the optimisation problem it defines in the manipulation-stage
of the game. Doing so does not change the incentives to exert effort, and we
therefore deduce that the optimal contract is manipulation-proof. However,
manipulation-proofness is not obtained because a manipulation is a "bad thing"
\textit{per see}, but simply because there are no losses in perfectly
preventing it.\\

Some readers might have further interest in the shape of the optimal contract.
We conclude this section with a corollary that helps characterize it
further. Since any continuous, non-negative and non-decreasing function can be
written as a maximum we deduce:
\begin{corollary}\label{generalizeddebt}Let $g$ satisfy Assumption 3. Then the
optimal contract can be written as a generalized debt contract
$$Y(x)=\max\{0,\alpha(x)x - d\} +w$$
where $d\geq 0$ is a threshold of debt, $w\geq 0$ is a flat wage and $\alpha(x)$
is a non-negative and continuous function with slope $\leq 1+r$. 
\end{corollary}

\subsection{Convex manipulation technologies}
In the last two sections we aimed to characterize the optimal contract in its
greatest generality and thus we tried to impose as few assumptions as possible.
Our goal now is to show that convex manipulation technologies sometimes lead to
contracts which induce manipulation in equilibrium. This is an existence
statement, which will allow us to make stronger assumptions in order to
highlight the mechanics at play.\\

Let us assume that the entrepreneur is risk-neutral and makes the
take-it or leave-it offer at the initial stage of the game. Let the
manipulation technology be \begin{align*}
    g(z) = \begin{cases}0 &\text{ if } z\leq 0 \\ \Tilde{g}(z) &\text{ if } z>0 \end{cases}
\end{align*}
for $\Tilde{g}$ a strictly convex function which is differentiable on $(0,M)$ and which satisfies
\begin{align}
    \lim_{z\downarrow 0}\Tilde{g}(z)=0 \quad \text{and} \quad \inf_{z>0} \left \{\frac{\partial \Tilde{g}(z)}{\partial z}\right \} = 1.\tag{Assumption 4} \label{Assumption 4}
\end{align}
It is important to highlight that Assumption 4 could be weakened by assuming that $\inf_{z>0} \left \{\frac{\partial \Tilde{g}(z)}{\partial z}\right \} \geq 1$, but this would only make the proof more tedious without gaining further insights. Finally, let us assume that the financier never agrees to give the entrepreneur more than the (state-by-state) profit of the business so that
$Y\leq X$.\\

The optimal contract solves the following optimisation problem
\begin{align}
\sup_{Y \in B_+(\mathcal{B}([0,M])), \overline{e} \in E} &\int Y(X(s,\overline{e}) +z(s)) -g(z(s)) d\mathbb{P}-c(\overline{e})  \tag{Problem Entrepreneur} \label{Problem MLRP}\\
  s.t.\, & \,0 \leq Y \leq X \tag{Feasibility} \label{FeasibilityMLRP}\\
       &\, \int X(s,\overline{e})+ z(s) - Y(X(s,\overline{e})+z(s))) d \mathbb{P} \geq (1+r)Q  \tag{IR} \label{IRMLRP}\\
       &\, \overline{e}\in \arg \max_e \left \{\int Y(X(s,e) +z(s)) -g(z(s))d\mathbb{P}  -c(e)\right \}\tag{IC} \label{ICMLRP}\\
       \forall s \in S,\, &\quad z(s) \in  \arg \max_z \{Y(X(s,\overline{e})+ z) -g(z) \} \tag{IIC} \label{IICMLRP}
\end{align}
Our current goal is to show that probability distributions and
manipulation technologies exist for which the optimal contract entails manipulations
in equilibrium. We do so by using bonus contracts, which define a
partition $$\mathcal{M}=\{[0,a),[a,b),[b,M]\}$$
of $[0,M]$ for which manipulations will be restricted to the middle interval
$[a,b)$. Intuitively, these intervals correspond to the three states $x_l,x_m$
and $x_h$ of the example of Section 1.2. We will be done by showing that 
situations where the probability of middle interval is small and the bonus
contract implements a strictly higher level of effort than the best-manipulation
proof contract can exist simultaneously. \\

First we claim that the constraint \eqref{IRMLRP} is binding at any solution
$Y$ of \ref{Problem MLRP}. This claim is standard and we do not prove it in the
main text. Let $e_{MP}$ be the highest level of effort which is implementable with a
manipulation-proof contract and let $Y^{MP}$ implement $e_{MP}$. We want to
know if we can find an alternative contract $Y$ and a level of effort $e_Y$
such that simultaneously $e_Y$ cannot be implemented with a manipulation-proof
contract and \begin{align*}
   \int Y(X(s,e_Y) +z(s)) -g(z(s)) d\mathbb{P}-c(e_Y) > \int Y^{MP}(X(s,e_{MP}))d\mathbb{P}-c(e_{MP}).
\end{align*}
\ref{Assumption 4} guarantees that every manipulation-proof
contract must be continuous and have a slope $\leq 1$. The manipulation-proof
contract which implement the highest level of effort is thus a simple debt
contract representable by the function \begin{align*}
    Y^{MP}(x)=\max\{0,x-d\}
\end{align*}
for $d\in (0,M)$ solving constraint \eqref{IRMLRP} at equality.\\

We want to show that there exists a bonus contract $Y^{Bonus}$ which dominates
the contract $Y^{MP}$. Thus, consider the contract \begin{align*}
    Y^{Bonus}(x)=\begin{cases} 0 &\text{ if } x<d' \\ x-\beta &\text{ if } x\geq d' \end{cases}
\end{align*}
for $0<\beta<d'$ and $d<d'<M$. The value $b=\beta - d'>0$ is the amount of
bonus the entrepreneur keeps upon a realization of profit greater than $d'$. Since $\Tilde{g}$ is strictly convex the contract $Y^{Bonus}$ defines a partition 
$$\mathcal{M}_{\Tilde{g}}=\{[0,d'-\Tilde{g}^{-1}(b)),[d'-\Tilde{g}^{-1}(b),d'),[d',M]\}$$
for which there is manipulation in the middle interval $[d'-\Tilde{g}^{-1}(b),d')$.
That is, the function
\begin{align*}
     z(x) = \begin{cases}0 &\text{ if } x \in \mathcal{M}_g \setminus [d'-\Tilde{g}^{-1}(b),d') \\ d'-x &\text{ if } x \in [d'-\Tilde{g}^{-1}(b),d')\quad \end{cases}
\end{align*}
is a selection of the optimal choice correspondence $\sigma(Y,x)$. As mentioned, these three intervals intuitively correspond to the three states
$x_l,\, x_m$ and $x_h$ we had in Section 1.2. Thus, it suffices to show that we can
find situations where \begin{align} \mathbb{P}[X(e)\in
  [d'-\Tilde{g}^{-1}(b),d')\vert e]\rightarrow 0, \label{convergence}
\end{align}
a convergence which intuitively corresponds to the limit $q\downarrow 0$ in the
example of Section 1.2. We can show this by finding a "sequence of increasingly
steeper functions" $\Tilde{g}$ so that the interval $[d'-\Tilde{g}^{-1}(b),d')$
converges to the singleton $\{d'\}$. The assumption that the family
$(X(e))_{e\in E}$ consists exclusively of continuous random variables then
guarantees the convergence in \eqref{convergence}. If $Y^{Bonus}$ implements a higher level of effort than $Y^{MP}$ and if the
effort is "productive enough" then we have shown that the former contract
dominates the latter. The notion of an effort level being "productive enough"
is elusive. The MLRP is enough for the argument above to be correct, although the
example in Section 1.2 show that it is not necessary.

\begin{theorem}\label{MLRP_and_manipulation}
Let the family $(X(e))_{e\in E}$ be ordered in the likelihood ratio and let
$e_{MP}<e_{2nd}$, where $e_{MP}$ is the highest level of effort implementable
with a manipulation proof contract and $e_{2nd}$ is the highest level of
effort when there is only ex-ante moral hazard.\\

Then we can always find a manipulation technology $g$ satisfying assumption 4
for which the solution to problem \eqref{Problem MLRP} entails manipulation in
equilibrium: there exists profit realizations $x\in [0,M]$ such that $z(x)>0$.
\end{theorem}
\section{Discussion}
   The possibility that the optimal contract entails manipulation in equilibrium
is sensitive to the interplay between the manipulation technology and the
stochastic output and thus to the assumptions we have made.
It is worth taking a closer look at the proof of Theorem \ref{MLRP_and_manipulation}
to better understand this sensitivity.  Theorem 8 relies on two key moving parts, the manipulation technology and the
distribution of output. A thorough understanding of both of these moving parts is useful to
the interpretation of the model. First, we want to emphasize that the assumptions we made about the manipulation
technologies are quite strong and were in fact binding our hands. Incidentally, we will show that
Theorem \ref{MLRP_and_manipulation} is more general than a first glance can tell.
\\

The proof of Theorem \ref{MLRP_and_manipulation} relies on comparing the best
manipulation-proof contract to a slightly modified version of itself. Intuitively,
this is a variational argument, as the bonus contract we consider is
essentially the best manipulation-proof contract to which we added an upward
jump at a well-chosen point. If the new contract implements a higher level of
effort and keeps the expected loss of manipulation low then we are done. We assumed a
manipulation technology for which the cost of small upward manipulation is
"large", which implied that the best manipulation-proof contract still
incentivizes working hard. However, the literature also considers technologies
for which the cost of small upward manipulations is "low", for instance by assuming that
\begin{align*}
    g(z)=g(\overline{x}-x)=(\overline{x}-x)^2
\end{align*}
where $x$ is once again the realized profit and $\overline{x}$ is the profit
as declared by the entrepreneur. Such type of manipulation technology is used both in the costly state falsification literature
\citep{crocker2007economics} and the earnings management literature \citep{sun2014executive}, where
it is interpreted as a situation where the manager can manipulate the
firm's accounting profit. Taking derivative around $z=0$ we see that
\begin{align*}
    \left. \frac{\partial g(z)}{\partial z}\right\vert_{z=0}=0 = \inf_{z>0}\left\{\frac{\partial g(z)}{\partial z}\right\}
\end{align*}
and any upward sloping contract induces manipulations. With such
manipulation technology it is impossible to incentivize working hard with a
manipulation-proof contract. In other words, the optimal contract \textit{always}
entails manipulation in equilibrium when the cost of small lies is low and
incentivizing hard work is valuable.\\

Which brings us to the second key moving part of the proof. As mentioned, the bonus
contract can be thought of as a local variation of the best manipulation-proof
contract, the debt contract. This local change to the contract implements a higher level of effort
if the effort "moves enough probability weight from the left to the right" of
the distribution. By definition the MLRP does precisely that, and is thus
essentially sufficient to show that this local change improves on the original
contract. However, the MLRP condition is not necessary for such perturbation argument to
be \textit{globally} true. Indeed, many distributions that do not satisfy
the MLRP  still exhibit the property that a well-chosen bonus contract
implements a higher level of effort than a debt contract. For instance, many
"U-shaped" distributions have this property, which is the intuition
that let us build the example in Section 1.2.
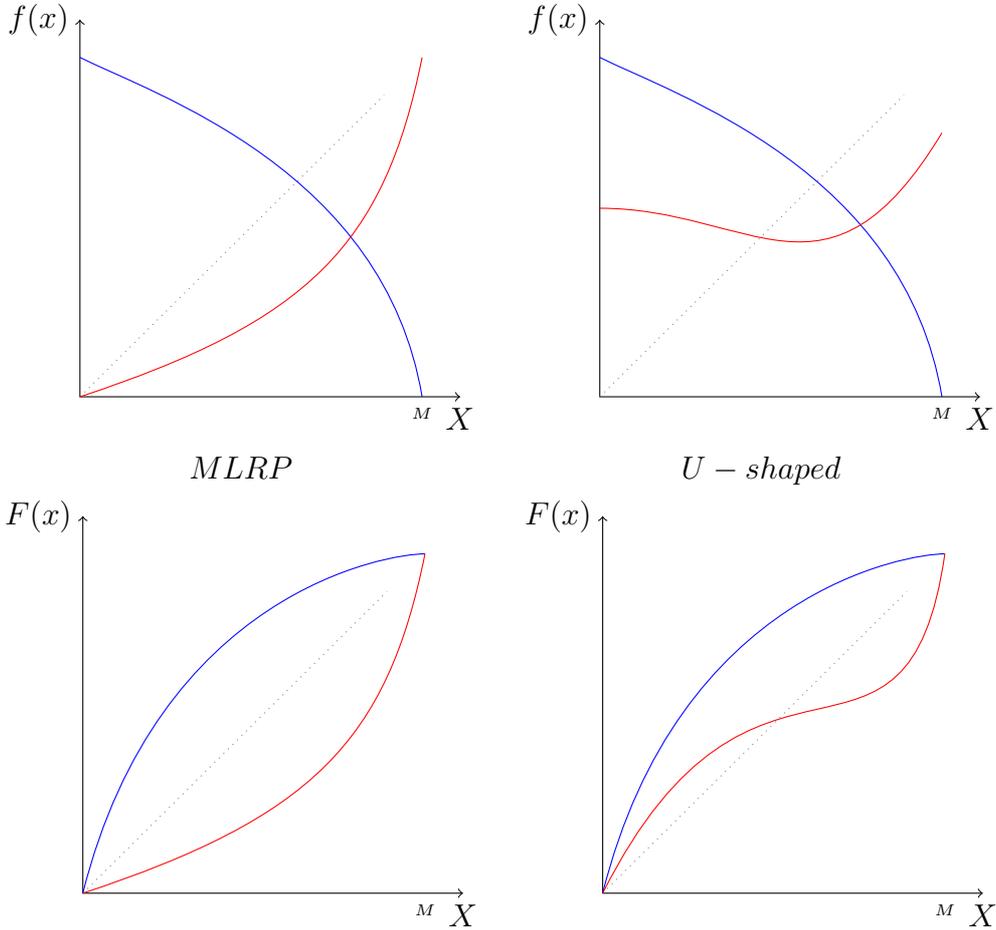
\begin{figure}[!ht]
    \centering
    \caption{MLRP \& U-shaped distributions}
    $\begin{array}{cc}
MLRP & U-shaped\\
\begin{tikzpicture}
\draw[<->] (0,5) -- (0,0) -- (5,0);
\draw[gray,dotted] (0,0) -- (4,4);
\node[below] at (5,0) {$X$};
\node[below] at (4.5,0) {\tiny{$M$}};
\node[left] at (0,5) {$f(x)$};
\draw [blue] (0,4.5) .. controls (1,4) and (4,3).. (4.5,0);
\draw [red] (0,0) .. controls (3,1) and (4,2).. (4.5,4.5);
\end{tikzpicture}  

&
\begin{tikzpicture}
\draw[<->] (0,5) -- (0,0) -- (5,0);
\draw[gray,dotted] (0,0) -- (4,4);
\node[below] at (5,0) {$X$};
\node[below] at (4.5,0) {\tiny{$M$}};
\node[left] at (0,5) {$f(x)$};
\draw [blue] (0,4.5) .. controls (1,4) and (4,3).. (4.5,0);
\draw [red] (0,2.5) .. controls (2,2.5) and (3,1).. (4.5,3.5);
\end{tikzpicture}  
\\
MLRP & U-shaped\\
\begin{tikzpicture}
\draw[<->] (0,5) -- (0,0) -- (5,0);
\draw[gray,dotted] (0,0) -- (4,4);
\node[below] at (5,0) {$X$};
\node[below] at (4.5,0) {\tiny{$M$}};
\node[left] at (0,5) {$F(x)$};
\draw [blue] (0,0) .. controls (1,4) and (4,4.5).. (4.5,4.5);
\draw [red] (0,0) .. controls (3,1) and (4,2).. (4.5,4.5);
\end{tikzpicture}  

&
\begin{tikzpicture}
\draw[<->] (0,5) -- (0,0) -- (5,0);
\draw[gray,dotted] (0,0) -- (4,4);
\node[below] at (5,0) {$X$};
\node[below] at (4.5,0) {\tiny{$M$}};
\node[left] at (0,5) {$F(x)$};
\draw [blue] (0,0) .. controls (1,4) and (4,4.5).. (4.5,4.5);
\draw [red] (0,0) .. controls (2,4) and (4,1).. (4.5,4.5);
\end{tikzpicture}  

\end{array}$
\end{figure}

We would like to conclude by explaining our own interpretation of our results,
an interpretation with which the reader may very-well disagree. Piecing together
the observations made above we consider that manipulation is often a
necessary evil. In our model's restricted world there are many situations where
the optimal contract induces acceptable manipulations in equilibrium.
If the model of \citet{crocker2007economics}
theoretically links the growth of performance-based executive compensation to
the explosion of accounting scandals of the early twenty-first century, ours
suggest that such theoretical link is not restricted to high-level executives,
as contracts with high-powered incentives are the staples of our modern
economy.\\

Assessing the scale of these acceptable manipulations remains an empirical question. One to
which we, the author, are skeptical can ever be answered precisely. 
The literature on earnings management consistently observes a
positive correlation between CEOs' incentive pays and earnings management.
Articles like \citet{sun2014executive} use simplified manipulation models to
argue that this correlation is likely to be driven by optimal contracting and
does not reflect inefficiencies in the market, further evidence that acceptable
manipulations exist.\\

We do not believe that more could be done. Our fundamental objection is one of
logical consistency, as assessing precisely the losses due to acceptable manipulations would require
that the econometrician observes both hidden actions of exerting effort and
profit manipulation. However, we postulate that the Principal cannot observe
these actions, as such observation would preclude moral hazard. In other words, evaluating
empirically such phenomena with any precision would require for the
econometrician to be better informed than the contracting parties, an
assumption which is hardly tenable in any situation we can think of.

\section*{Conclusion}
    
The literature on ex-post moral hazard is well established, dating at least to
the costly state verification model of \cite{townsend1979optimal}. However, the
subsequent literature considers many different definitions for a manipulation,
and the conclusions for each particular model is highly sensitive to the
assumptions made about the manipulation technologies. The recent literature highlights a trade-off between the provision of
incentives to work hard and the prevention of manipulation. The importance of
this trade-off is supported by the empirical literature on earnings
management, which consistently observes a positive correlation between CEOs'
incentive pay and earnings management.\\

Despite many theoretical models and empirical evidence pointing to the
existence of such trade-offs, no previous article provides a set of general
conditions under which the optimal contract entails acceptable manipulations in
equilibrium. This state of knowledge is unfortunate given the strong normative
implications of some models, which implicitly imply that unethical behaviours
such as fraud are a normal part of a well-functioning economy. This article sheds light on the interplay between the manipulation technology
and the productivity of effort. The optimal contract is non-decreasing whenever
the agent can burn the business' profit. This is because burning money is
unambiguously wasteful and no gains in incentives can be made by allowing it.\\

The optimal contract is always manipulation-proof when the manipulation
technology is linear. This feature is entirely driven by the linearity
assumption, which guarantees that any contract can be replaced by the value
function of the optimisation problem it defines in the manipulation stage of
the game without changing incentives. In other words, the reason why the
optimal contract prevents manipulation is not because a manipulation is
"bad"\textit{ per see}, but because there are no losses in doing so when the
technology is linear. This is not true with convex manipulation technologies.\\

When the manipulation technology is convex then the optimal contract sometimes
entails acceptable manipulations in equilibrium. This feature depends on the specificity of
the interplay between the manipulation technology and the "productivity of
effort". Intuitively, when working hard is productive enough to be worth
rewarding, then upward manipulations are justified, provided that the expected losses to manipulation stays low. Bonus
contracts have the desirable property of incentivizing hard work while
maintaining the expected losses to acceptable manipulations low.\\

A mathematical definition of "productive enough" effort is elusive.
We have shown that the monotone likelihood ratio is
enough to prove the existence of optimal contracts which entail
manipulations in equilibrium. It is not, however, necessary, as bonus contracts can
incentivize a high level of effort for many other types of distributions.
However, our results still suggest that the monotone likelihood ratio is a
stronger assumption than previously thought, as it is essentially the type of
assumption which justifies acceptable manipulations. That is, it is precisely
the type of assumption for which it is true that unethical behaviours such as
fraud are a normal part of a well-functioning economy.

\nocite{*}

\break 
\bibliographystyle{aer}
\bibliography{bibliography.bib}

\break 
\appendix
\section{Omitted definitions}
   
We collect standard results on stochastic orders. We mainly follow the treatment of \citet{shaked2007stochastic} but we also incorporate some results known in the literature. We assume throughout that every random variable has support $\chi \subset[0,M]$ for $0<M<\infty.$\\

Let $X,Y$ be two random variables. We say that\textbf{ $X$ is smaller than $Y$ in the usual stochastic order}, denoted by $X\leq_{FOSD} Y$, if 
\begin{align}
    \mathbb{P}[X>x]\leq \mathbb{P}[Y>x] \quad \text{for all }x\in \chi. \tag{FOSD} \label{FOSD}
\end{align}
Condition \eqref{FOSD} is often called \textbf{first-order stochastic dominance}. Let $F$ and $G$ denote the cumulative distribution function of $X$ and $Y$ respectively. It holds that $X\leq_{FOSD}Y$ if and only if
\begin{align*}
    G(x)\leq F(x) \text{ for all }x \in \chi \text{ with strict inequality for some }x.
\end{align*}
Accordingly, we write $F\leq_{FOSD}G$ to denote $X\leq_{FOSD}Y$ when it is not ambiguous. Let $(X(\theta))_{\theta \in \Theta}$ be a family of random variable with parameters $\theta \in \Theta \subset \mathbb{R}$. Let $(F(x\vert \theta))_{\theta \in \Theta}$ be their corresponding conditional cumulative distribution functions and assume that $F(\cdot \vert \theta)$ is differentiable in $\theta$. Then $(F(x\vert \theta))_{\theta \in \Theta}$ satisfy FOSD in $\theta$, $\theta\leq \theta' \imp F(x\vert \theta)\leq_{FOSD}F(x\vert \theta')$, if
\begin{align*}
    F_\theta(x \vert \theta) \leq 0 \text{ for all } x \in \chi \text{ with strict inequality for some } x.
\end{align*} 

Let $X,Y$ be random variables and let $f,g$ be their corresponding density. Let $$L(x):=\frac{g(x)}{f(x)}$$ be their likelihood ratio. We say that \textbf{X is smaller than Y in the likelihood ratio order}, denoted by $X\leq_{LR}Y$, if 
\begin{align}
\frac{\partial L(x)}{\partial x} \geq 0 \text{ for all } x\in \chi \tag{MLRP} \label{MLRP} 
\end{align}
where $a/0:=\infty$ whenever $a>0$. Condition \eqref{MLRP} is sometimes called the \textit{monotone likelihood ratio property}, and is equivalent to the condition that 
\begin{align*}
    f(x)g(y)\geq f(y)g(x) \text{ for all } x\leq y.
\end{align*}
Integrating the previous equation over $x\in A$ and $y \in B$ for $A,B$ measurable subsets of $\chi$ we obtain the following equivalent condition: 
\begin{align*}
\mathbb{P}[X\in A] \mathbb{P}[Y\in B] \geq \mathbb{P}[X\in B] \mathbb{P}[Y\in A]  \text{ for all measurable sets $A,B$ such that $A\leq B$}
\end{align*}
where $A\leq B$ means $x\in A$ and $y\in B$ implies $x\leq y$. This last condition is interesting because it does not involve densities and applies uniformly to continuous, discrete or mixed distributions.\\

Let $X$ and $Y$ have full support and denote by $F$ and $G$ their respective cumulative distribution functions. Then 
\begin{align*}
X\leq_{LR}Y \iff G/F \text{ is convex}.
\end{align*}
Let $(X(\theta))_{\theta \in \Theta}$ be a family of random variables with full support and let $(f(x\vert \theta))_{\theta \in \Theta}$ be their corresponding conditional densities. Assume that $f(\cdot \vert \theta)$ is differentiable in $\theta$. Then $(f(x\vert \theta))_{\theta \in \Theta}$ satisfies the MLRP in $\theta$, $\theta\leq \theta' \imp X( \theta)\leq_{LR}X( \theta')$, if
\begin{align*}
  f(x\vert\theta)f(y\vert \theta') \geq f(x\vert \theta') f(y\vert \theta) \text{ whenever $x>y $ and $\theta' >\theta$}.
\end{align*} 
The previous condition is equivalent to 
\begin{align*}
    \frac{\partial}{\partial x}\left[\frac{f_\theta(x\vert \theta')}{f(x\vert \theta')}\right]\geq 0 \text{ for all } \theta'\in \Theta \text{ and for all } x\in \chi.
\end{align*}
We say that $(f(x\vert \theta))_{\theta \in \Theta}$ satisfies the strict MLRP in $\theta$ if the previous inequality is strict. Alternatively, the strict MLRP states that for every $ \theta < \theta'$ it is
\begin{align*}
  f(x\vert\theta)f(y\vert \theta') > f(x\vert \theta') f(y\vert \theta) \text{ whenever $x>y $ and $\theta' >\theta$}.
\end{align*}
Of course the strict MLRP implies that $F(\theta')/ F(\theta)$ is strictly convex. Finally, note that 
    \begin{align*}
        X\leq_{LR} Y \implies X\leq_{FOSD} Y
    \end{align*}
but the converse is not generally true unless $\vert \chi \vert=2$. 
   
\break
\section{Ommitted proofs}
    
\noindent \textbf{Proof of Lemma 2:} Lemma 2 follows from Theorem 3 of \citet{Lauzier2019positioningmaths}. We sketch an alternative, simpler proof. We first aim to prove that $V$ is continuous. Suppose, by the way of contradiction, that $V(x_{s,e},Y)$ is discontinuous somewhere in the interior of its domain, i.e. there exists a $x_0\in(0,M)$ such $V(x_0^-,Y)\neq V(x_0^+;Y)$. By Lemma 1, $V$ is non-decreasing and so it must be the case that $V(x_0^-;Y)<V(x_0^+;Y)$. Hence, by the continuity of $g$ there exists a $\Tilde{x}$ such that $\Tilde{x}- x_0^+ = \Tilde{z}\in \sigma(Y,x_0^+)$ and
$$\limsup_{x\rightarrow x_0^-} Y(x+\Tilde{z})-g(\tilde{z})= V(x_0^+;Y)$$ and $\Tilde{z}$ dominates every $z\in \sigma(Y,x_0^-)$, an absurd. Continuity of $V$ at endpoints, 
$$\lim_{x \downarrow 0} V(x;Y)= V(0;Y) \quad \text{and} \quad  \lim_{x \uparrow M}V(x;Y)= V(M;Y),$$ follows from Assumption 2. Notice that since $Y$ is bounded $V$ also is, and thus $V$ is Lipschitz. $\blacksquare$\\

\noindent \textbf{Proof of Lemma 5:} Lemma 5 follows from Corollary 4 of \citet{Lauzier2019positioningmaths}. We sketch an alternative, simpler proof. Since $V$ is Lipschitz, by Rademacher's theorem it is almost everywhere differentiable, i.e. the derivative $\frac{\partial V(x;Y)}{\partial x}$ is almost everywhere well-defined. The inequality $0\leq \frac{\partial V(x;Y)}{\partial x}$ follows immediately from Lemma \ref{monotonicityVF}. Suppose, by the way of contradiction, that there exists a $x_0\in (0,M)$ such that $$\left. \frac{\partial V(x;Y)}{\partial x}\right\vert_{x=x_0}>1+r.$$
Let $z\in \sigma (Y,x_0)$ and notice that $z\geq 0$ by Lemma \ref{monotonicityVF}. Let $\varepsilon>0$ and observe that our contradiction hypothesis implies that for $\varepsilon$ small it is
$$Y(x_0-\varepsilon +z) - g(x_0-\varepsilon +z)= Y(x_0-\varepsilon +z) - (1+r)(x_0-\varepsilon +z)> V(x_0-\varepsilon; Y),$$
an absurd. $\blacksquare$\\

\noindent \textbf{Proof of Theorem 6:} By Theorem \ref{MonotonicityY} the optimal contract is non-decreasing. Suppose by contraposition that the contract $Y$ induces manipulation in equilibrium and let $e_Y$ be a level of effort implemented by $Y$. Since $Y$ is monotonic the contraposition assumption states that there exists some $x_{s,e_Y}\in [0,M]$ for which every $z\in \sigma(Y,x_{s,e})$ are such that $z>0$. Let $V(x_{s,e_Y};Y)$ be the value function of the manipulation stage of the game and let $U\subset [0,M]$ be the set of $x_{s,e_Y}\in [0,M]$ for which the contract Y induces manipulation in equilibrium. By Lemma 5, $V(x_{s,e_Y};Y)$ is continuous and has slope $\leq 1+r$, with equality on $U$. Consider the alternative contract $Y$ defined by the value function $V(x_{s,e_Y};Y)$, i.e. $\overline{Y}=V(x_{s,e_Y};Y).$ Since $g$ is linear it suffice to prove that $\overline{Y}$ is manipulation-proof to obtain that $\overline{Y}$ implements $e_Y$ and dominates $Y$, similarly to the proof of Theorem \ref{MonotonicityY}.\\

Redefine the optimal manipulation correspondence as
\begin{align*}
    &\sigma(Y,x)=\arg\max_{z}\{Y(x+z)-g(z)\} \quad \text{and}\\
    &\sigma(\overline{Y},x)=\arg\max_{z}\{\overline{Y}(x+z)-g(z)\}.
\end{align*}
Notice that since $Y$ is manipulation-proof on $[0,M]\setminus U$ the two contracts are equal on this set and thus $\overline{Y}$ is also manipulation-poof on $[0,M]\setminus U$. Let $x\in U$ be given. We want to show that $0\in \sigma(\overline{Y},x)$. By construction to every $z\in \sigma(Y,x)$ it is $\overline{Y}(x)= Y(x+z)-(1+r)z$. Suppose by contradiction that $0 \notin \sigma(\overline{Y},x)$. Since $\overline{Y}$ is monotonic this implies that every $z'\in \sigma(\overline{Y},x)$ are such that $z'>0$. However, this assumption implies that \begin{align*}
    \overline{Y}(x+ z')-(1+r)z' & > \overline{Y}(x) \quad \iff\\
    [Y(x+z') -(1+r)z'] - (1+r)z' & > \overline{Y}(x) \quad \iff \\
    Y(x+z') - 2(1+r)z' & > \overline{Y}(x).
\end{align*}
If $z'\in \sigma(Y,x)$ then the previous equality becomes
$$Y(x+z') - 2(1+r)z'  > \overline{Y}(x)=Y(x+z')-(1+r)z',$$
an absurd given that $z'>0$ and $r>0$. If $z'\notin \sigma (Y,x)$ then there exists a $\Tilde{z}\in \sigma (Y,x)$, $\Tilde{z}\neq z'$, such that simultaneously
$$Y(x+\Tilde{z})-(1+r)\Tilde{z} > Y(x+z')-(1+r)z'$$
and
$$Y(x+\Tilde{z})-(1+r)\Tilde{z}= \overline{Y}(x)<\overline{Y}(x+ z')-(1+r)z',$$
another absurd. Thus, $0\in \sigma(\overline{Y},x)$ and we are done. $\blacksquare$\\

\noindent \textbf{Proof of Theorem 8:}
We begin with a few preliminary claims.
\begin{claim} The Individual Rationality constraint \ref{IRMLRP} must be binding at any solution of problem \ref{Problem MLRP}.\end{claim}
\noindent\textbf{Proof of the Claim:} Let $e_Y$ be the level of effort implemented by contract $Y$. The claim follows immediately by contraposition observing that if  $$\int X(e_Y) - Y(X(e_Y))d\mathbb{P} > (1+r)Q$$ then there exists an alternative contract $\Tilde{Y}\neq Y$ which implements effort $e_{\Tilde{Y}}\geq e_{Y}$, that is feasible because  
$$\int X(e_Y) - Y(X(e_Y))d\mathbb{P} > \int X(e_{\Tilde{Y}}) - \Tilde{Y}(X(e_{\Tilde{Y}})d\mathbb{P}\geq  (1+r)Q$$
and that strictly dominates $Y$ because
$$\int \Tilde{Y}(X(e_{\Tilde{Y}})d\mathbb{P}> \int Y(X(e_Y))d\mathbb{P}. \quad \quad \square$$

\begin{claim} The best manipulation-proof contract is a debt contract,  the function $$Y^{MP}(x)=\max\{0, x-d\}$$
for $d\in (0,m)$ satisfying
$$\int X(s,e_{MP})- \max\{0, X(s,e_{MP})-d\}d\mathbb{P}= (1+r)Q,$$
where $e_{MP}$ is the level of effort implemented by $Y^{MP}$.\end{claim}

\noindent \textbf{Proof of the Claim:} The feasibility constraint states that $0\leq Y \leq X$ and thus it is $Y^{MP}(0)=0$. By theorem \ref{MonotonicityY} the contract $Y^{MP}(x)$ is a non-decreasing function and since $e_{MP}<e_{2nd}$ the best manipulation-proof contract implements the highest possible effort level. By  \ref{Assumption 4} it is 
$$\inf_{z>0}\left \{\frac{\partial g(z)}{\partial z}\right \} = 1$$
and $Y^{MP}$ must be continuous and with a slope $\leq 1$ in order to prevent manipulations. Set $$Y^{MP}(x)=\max\{0, x-d\}$$ for $d>0$ making constraint \ref{IRMLRP} an equality. Taking derivative we have
$$\left . \frac{\partial Y(x)}{\partial x}\right \vert_{x\in (0,M)\setminus\{d\}}= \begin{cases}0 &\text{if}\, \, x<d\\ 1 &\text{if} \,\, x>d\end{cases}$$
and the monotone likelihood ratio property guarantees that $Y^{MP}$ is the manipulation-proof contract that implements the highest possible effort level. $\square$\\

We want to show that there exists a manipulation technology for which there is a pair $(Y,e_Y)$ such that $Y$ implements $e_Y$ and $Y$ strictly dominates $Y^{MP}$, i.e
\begin{align}
\int Y(X(s,e_Y)+z(s)-g(z(s))d\mathbb{P}-c(e_Y)> \int Y^{MP}(X(s,e_{MP})d\mathbb{P} - c(e_{MP}) \tag{Domination} \label{domination}
\end{align}
for $z(s)\in \sigma(Y,x_{s,e_Y})$
a selection. Rearranging we obtain
\begin{align*}
    \mathbb{E}[Y(X(e_Y)]-\mathbb{E}[Y^{MP}(X(e_{MP})] - [c(e_Y)-c(e_{MP})]> \mathbb{E}[g(z(s))].
\end{align*}
The assumption that $e_{MP}<e_{2nd}$ guarantees that there exists effort levels $e_{\Tilde{Y}}\in (e_{MP}, e_{2nd}]$ for which we can find a contract $\Tilde{Y}$ which implements $e_{\Tilde{Y}}$ and is such that
\begin{align*}
    \mathbb{E}[\Tilde{Y}(X(s,e_{\Tilde{Y}})]-\mathbb{E}[Y^{MP}(X(s,e_{MP})] - [c(e_{\Tilde{Y}})-c(e_{MP})]>0.
\end{align*}
Thus, we will be done if we can find a pair $(Y,e_Y)$ such that $e_Y\in (e_{MP}, e_{2nd}]$ and 
$$\mathbb{E}[g(z(s))]\rightarrow 0,$$
a convergence which we will define precisely below.\\

Consider the following bonus contract:
\begin{align*}
    Y^{Bonus}(x)=\begin{cases} 0 &\text{ if } x<d' \\ x-\beta &\text{ if } x\geq d' \end{cases}
\end{align*}
for $0<\beta<d'$ and $d<d'<M$. By Theorem 2 of \citet{Lauzier2019positioningmaths} for every given $g$ satisfying  \ref{Assumption 4} there exists a partition $$\mathcal{M}_{\Tilde{g}}=\{[0,d'-\Tilde{g}^{-1}(b)),[d'-\Tilde{g}^{-1}(b),d'),[d',M]\}$$
for which the function 
\begin{align*}
     z(x) = \begin{cases}0 &\text{ if } x \in \mathcal{M}_g \setminus [d'-\Tilde{g}^{-1}(b),d') \\ d'-x &\text{ if } x \in [d'-\Tilde{g}^{-1}(b),d')\quad \end{cases}
\end{align*}
is a selection of the optimal choice correspondence 
$$\sigma(Y^{Bonus},x)=\arg\max_{z}\{Y^{Bonus}(x+z)-g(z)\}.$$
There exists a net of functions $(g_\gamma)_{\gamma \in \Gamma}$ such that \begin{enumerate}
    \item for every $\gamma<+\infty$ the function $g_\gamma$ satisfies Assumption 4;
    \item the net $(g_\gamma)_{\gamma \in \Gamma}$ consists of increasingly steeper functions: if $\gamma'>\gamma$ then for every $z$ it is $g_{\gamma'}(z)\geq g_{\gamma}(z)$, with strict inequality for some $z$ and
    \item it is 
    \begin{align*}
        \lim_{\gamma\rightarrow +\infty} g_{\gamma}(z)=g_{\infty}(z) = \begin{cases} 0 &\text{if} \, \, z\leq 0 \\ +\infty & \text{if}\, \,z>0\end{cases}.
    \end{align*}
\end{enumerate}
The middle interval $[d'-\Tilde{g}_\gamma^{-1}(b),d')$ converges to the singleton $\{d'\}$, i.e
$$[d'-\Tilde{g}_\gamma^{-1}(b),d')\overset{\gamma \rightarrow +\infty}{\longrightarrow} \{d'\}.$$
Since the random variables $X(e)$ are continuous it holds for every $e\in E$ that
\begin{align*} \mathbb{P}[X(e)\in
  [d'-\Tilde{g}_\gamma^{-1}(b),d')\vert e] \overset{\gamma \rightarrow +\infty}{\longrightarrow} 0
\end{align*}
and thus 
\begin{align*}
  \mathbb{E}[g_\gamma(z(s))]  \overset{\gamma \rightarrow +\infty}{\longrightarrow} 0.
\end{align*}
As $\gamma \rightarrow +\infty$ the Incentive Compatibility constraint \ref{ICMLRP} converges to the Incentive Compatiblity constraint of the problem with ex-ante moral hazard only, i.e. to the Incentive Compatibility constraint of the problem
\begin{align*}
\sup_{Y \in B_+(\mathcal{B}([0,M])), \overline{e} \in E} &\int Y(X(s,\overline{e}) )) d\mathbb{P}-c(\overline{e})  \\
  s.t.\, & \,0 \leq Y \leq X  \\
       &\, \int X(s,\overline{e}) - Y(X(s,\overline{e}))) d \mathbb{P} \geq (1+r)Q \\
       &\, \overline{e}\in \arg \max_e \left \{\int Y(X(s,e)) d\mathbb{P}  -c(e)\right \}.
\end{align*}
Since the distribution satisfies the monotone likelihood ratio property we can define the bonus contract $Y^{Bonus}$ such that, in the limit when $\gamma \rightarrow +\infty$, the contract implements the effort level $e\in (e_{MP},e_{2nd}]$ and satisfies the constraint \ref{IRMLRP} at equality. Thus, the inequality (\ref{domination}) is satisfied in the limit as $\gamma \rightarrow +\infty$ and we are done. $\blacksquare$
    
    \break
\section{Solution concept}
    
This appendix expands on the solution concept we used in the main text and explains further some of the assumptions we made. We focus on the case where the financier makes the take-it or leave-it offer at the initial stage of the game.\\

Recall that the entrepreneur have to make two sequential choices after being presented with an offer; he first chooses a level of effort $e\in E$, then he observes the realisation $X(s,e)$ ("Nature moves") and finally he chooses a manipulation $z\in \mathbb{R}$. We would like to emphasis that this wording already implies that we are restricting our attention to equilibria in pure strategies. Now recall that the family $(X(e))_{e\in E}$ consists of random variables which all have common and full support $[0,M]$. We can thereby slightly abuse notation and write the optimal manipulation correspondence $\sigma(Y,e,s)$ as $\sigma(Y,x)$. In other words, our assumption guarantees that the optimal manipulation correspondence is a mapping $[0,M]\rightarrow \mathcal{B}([0,M])$.\\

The optimal manipulation correspondence is rarely single-valued, and not every actions which are payoff equivalent to the entrepreneur are equal. This is better seen by considering a flat part of a contract and noticing that the entrepreneur might be indifferent between burning the money or telling the truth, but that the former action hurts the financier while the latter does not. That is, the latter manipulation Pareto dominates the former. This motivate our focus on equilibria for which the entrepreneur takes the manipulation most favoured by the financier whenever indifferent, which is tantamount to focusing on equilibria for which the manipulation is Pareto efficient.\\

A similar problem arises for the choice of effort $e\in E$. Let the contract $Y$ be given and denote by $E^*(Y)$ the set of effort that maximize the expected payoff for the entrepreneur. The set $E^*(Y)$ we considered in the text does not need to be a singleton. In other words, a given contract does not necessarily implement only one level effort. This is because we worked with the weakest assumptions on the probability distribution as we can manage. Thereby we considered pure strategy equilibria where the entrepreneur chooses the highest level of effort whenever indifferent. This choice is motivated by our interpretation that working hard is a "good thing" and not by any  mathematical properties of the model.\\

This discussion incidentally sheds lights on our notion of sub-game perfectness. Our assumptions about the entrepreneur's actions at indifference guarantee that the financier's conjecture about the behaviour of the entrepreneur is correct in equilibrium. Without these assumptions there might exist equilibria where this conjecture is incorrect. Formally let $Y$ be an offer and let $E^*(Y)=\{e_l,e_h\}\subset E$ for $e_l<e_h$. Suppose that the financier believes that the entrepreneur plays $e_h$ with probability 1, i.e. that his conjecture is $$\mu^F(e^*(Y))=\delta_{e_h},$$ where $\mu^F(e^*(Y))$ is the financier's conjecture about the entrepreneur's choice of effort given offer $Y$ and where $\delta$ is the Dirac measure. Since $e_h\in E^*(Y)$ there might exists equilibria where the financier offer $Y$ and wrongly believes that the entrepreneur takes the action $e_h$ while the entrepreneur truly takes the action $e_l$. In other words, without our assumptions on the entrepreneur's behaviour, we would need to be very careful about our definition of belief and sub-game perfectness.

\end{document}